\documentclass[aps,prb,amsmath,amssymb,preprint]{revtex4}
\usepackage{graphicx, color}

\begin{document}
\title{Interplay of hybrid and unhybrid quantum transport of interacting electron pairs through a short conduction channel}

\author{Danhong Huang$^{1}$, Godfrey Gumbs$^2$, Yonatan Abranyos$^{2}$, Michael Pepper$^{3,4}$ and Sanjeev Kumar$^{3,4}$}
\affiliation{
$^{1}$Air Force Research Laboratory, Space Vehicles
Directorate, Kirtland Air Force Base, New Mexico 87117, USA\\
$^{2}$Department of Physics and Astronomy, Hunter College of the
City University of New York, 695 Park Avenue, New York, New York 10065, USA\\
$^{3}$Department of Electronic and Electrical Engineering, University College London, London, WC1E 7JE, United Kingdom\\
$^{4}$London Centre for Nanotechnology, 17-19 Gordon Street, London, WC1H 0AH, United Kingdom
}

\date{\today}

\begin{abstract}
For quantum ballistic transport of electrons through a short conduction channel, the role of Coulomb interaction
may significantly modify the energy levels of an electron pair at low temperatures as the channel becomes wide.
In this regime, the Coulomb effect on the orbital triplet and singlet electron-pair state is calculated and found to
lead to four split energy levels, including two hybrid and two unhybrid states.
Moreover, due to the interplay of hybrid and unhybrid Coulomb interactions between two electrons, our calculations reveal that
the ground pair-state will switch from one hybrid orbit-triplet state (strong confinement)
to the unhybrid orbit-singlet state (intermediate confinement) as the channel
width gradually increases and then back to the original hybrid orbit-triplet state (weak confinement), due to larger total spin of the orbit-singlet state,
as the channel width becomes larger than a threshold value.
This switching behavior leaves a footprint in the conductance as well as in the
diffusion thermoelectric power of electrons. Here, the predicted reoccurrence of the hybrid orbit-triplet state (spin-$0$ state) as a ground state
is shown to relate to the higher spin degeneracy of the spin-$1$ state as well as
to the strong Coulomb repulsion in the central region of the channel, which separates two electrons away and pushes them to different channel edges.
The conductance reoccurrence region expands from the weak to the intermediate confinement regime with increasing linear electron density.
\end{abstract}

\pacs{73.21.Hb,\,71.70.Ej,\,73.23.Ad,\,73.63.Nm}
\maketitle

\section{Introduction}
\label{sec:1}

For many years now, there has been a concerted effort to
understand the behavior of the conductance of quantum wires under variable
conditions of disorder, wire width and temperature for diffusive electron transport (for
example, see Refs.\,[\onlinecite{paper1,paper2,paper3,paper4,paper5,paper6,paper7,paper8}]).
For pure narrow samples of quantum wires
whose widths are a few nanometers, the conductance plateaus are
obtained as integer multiples of $2e^2/h$. It turns out
that since the kinetic energy dominates over the Coulomb
interaction in the limit of strong confinement, the conductance
plateaus at integer multiples of $2e^2/h$ may be adequately
accounted for in pure samples with the use of a single-particle
picture\,\cite{lyo5}. However, as the width of the wire is increased,
the Coulomb interaction between electrons plays more and more
of a role in determining the values of the conductance plateaus.
\medskip

The structural transition in a quasi-one-dimensional quantum-wire system was numerically predicted\,\cite{new9} as early as in 2004 with a rich phase diagram.
Later, a theoretical model for a split Wigner crystal into two chains (zigzag crystal) was proposed\,\cite{wire2} in 2007 (for a review, see Ref.\,[\onlinecite{wire4}]).
The effects of different pairwise repulsive interactions\,\cite{new6}, tunnel coupling of two parabolic channels\,\cite{new7}, different profiles of the confining channel\,\cite{new8},
and even a quantum-ring structure\,\cite{new5} or the the surface of a cylinder\,\cite{wire5}, on the continuous structural transitions of a Wigner crystal were further studied.
Similar structural transitions of a Wigner crystal to a zigzag crystal in an ion chain\,\,\cite{new1} and in quantum wires controlled by an external gate\,\cite{new2} were also explored.
In addition, the spin Peierls quantum phase transition in cold Coulomb crystals of trapped ions\,\cite{new11}, the spontaneous spin polarization due to the electron-electron interactions
under a bias-field control\,\cite{wire1,wire6}, and the phase diagram of zigzag Wigner crystals with spin coupling for two-, three and four-particle ring exchange processes\,\cite{wire3},
as well as melting of a quasi-one-dimensional Wigner crystal
observed from the nonlinear resistivity\,\cite{new10} of electrons confined in quasi-one dimensional channels formed on the surface of superfluid $^4$He, were reported.
\medskip

Despite the extensive theoretical studies on zigzag crystals for a long quantum wire, however,
in a recent experiment\,\cite{pepper} for a short conduction channel formed by split gates, one finds that,
as the top gate voltage is increased, the conductance for a wide wire at the interface of a
GaAs/AlGaAs heterostructure jumps from zero to $4e^2/h$,
bypassing the $2e^2/h$ plateau which is encountered for narrow
wires. For the picture of splitting into two rows, the evolution of the crossing or anti-crossing of energy levels is not fully understood, and the purpose
of this paper is to provide a microscopic theory which explains this observed phenomenon.
\medskip

Additionally, we predict that the
competition between the kinetic, direct Coulomb and quantum mechanical exchange
energy in wires of intermediate widths should lead to fundamental
differences from that obtained  in the two extreme limits of very narrow and very wide wires.
We have demonstrated that these differences may be traced to the nature
of the ground state as the wire width is varied.
\medskip

In related work, there have been several physical properties of the measured
conductance of quasi-one-dimensional quantum wires which have
been attributed to scattering from disorder potentials, the formation
of a quantum dot within the channel caused by the presence of an
impurity, as well as imperfections in the device geometry.\,\cite{paper5}
These imperfections may lead to deviations from integer multiples
of $2e^2/h$ for the values of the conductance plateaus or resonance structure
such as oscillations superimposed on the conductance
trace.\,\cite{paper4,half-step} Electron tunneling through the
quantum dot in the channel as well as interference effects due to
electron back-scattering from an impurity potential are believed to be responsible for these
deviations in the values of the conductance plateaus of narrow quantum
wires.\,\cite{paper4}
\medskip

In chemistry, hybridization is the well-known concept for mixing atomic orbitals into new hybrid ones (with different energies, shapes, etc., than the component atomic orbitals),
suitable for the pairing of electrons to form chemical bonds in valence bond theory.
Electronic orbital hybridization discussed in this paper means the mixing of orbitals of two interacting electron pairs to form a new ground and excited paired states\,\cite{pepper}.
The Coulomb interaction for electron pairs can be used for building up the hybrid states while the
pair ballistically passes along a one-dimensional conduction channel.
Our calculations reveal the role and the existence of these interacting hybrid states in quasi-one-dimensional quantum ballistic transports.
More importantly, we demonstrate in our work that the degree and significance of hybridization (Coulomb-induced level anticrossing) within such a structure may
be tuned independently by varying the channel confinement with the use of a top gate.
\medskip

In this paper, we confine our attention to a quasi-one-dimensional quantum wire
containing a low density of electrons. We concentrate our efforts
on calculating the lowest eigenstates for a pair of interacting
electrons since this sheds some light on the role played by
electron-electron interaction in determining the nature
of the ground state of a dilute electron system and consequently
the lowest quantum conductance.
The complicated pair tunneling process\,\cite{pair,pair1} will not be considered here since it does not lead to conductance plateaus observed in our experiment.
We show below that there is a range of values of wire widths where the two-electron transport
are hybrid by Coulomb interaction, and therefore, it is not possible to describe
the conductance in terms of a single-particle picture.

\section{Model}
\label{sec:2}

The eigenstates of a pair of interacting electrons under the
influence of a harmonic confining potential have been evaluated by
several authors.\,\cite{Wagner,Bryant}
In the paper by Wagner, et al.,\,\cite{Wagner} a quantum
dot with a symmetric harmonic oscillator potential to
confine the electrons was considered and it was noted that, consistent with Kohn's
theorem\,\cite{Bryant}, the Coulomb interaction affects only the relative
motion but not the center-of-mass properties. It was then demonstrated
with the use of perturbation theory that as the strength of an external
perpendicular magnetic field is increased, the ground state oscillates
between a spin-singlet and a spin-triplet mode. Bryant\,\cite{Bryant} showed
correlation effects between electrons depend on the size of
the boxes containing them. By solving the Schr\"odinger equation exactly
for a pair of interacting electrons, we demonstrate how correlations
may determine the ground state and give rise to quasi-particles which
participate in the transport processes.
\medskip

\begin{figure}[t]
\centering
\includegraphics[width=0.49\textwidth]{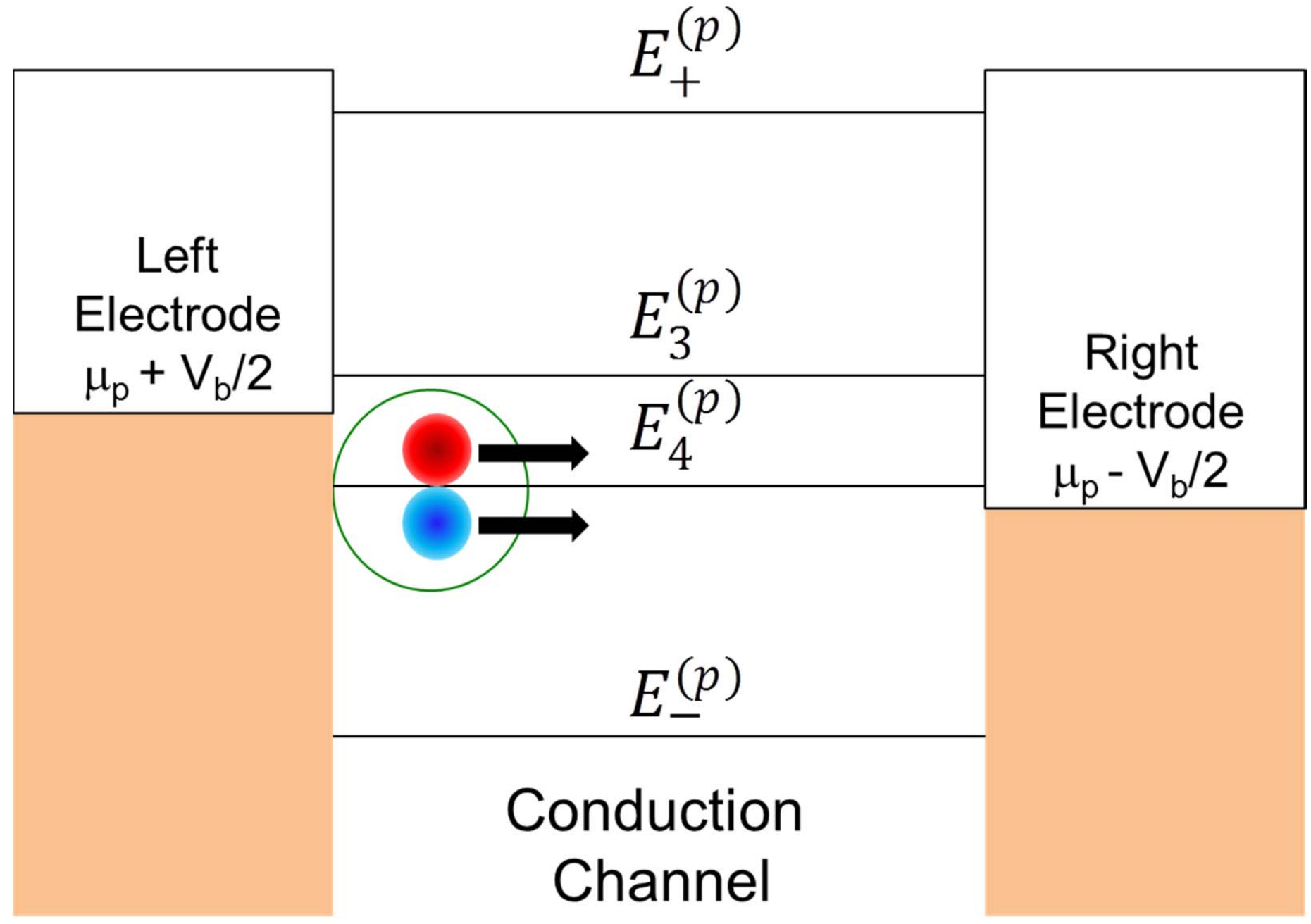}
\caption{(Color online) Schematic energy diagram of our model system, where an interacting electron pair is assumed to transport ballistically through a conduction channel,
where the labels for different energy levels are explained in the text.}
\label{FIG:0}
\end{figure}

If the scattering by either randomly distributed impurities and defects or by
phonons are neglected at low temperatures for high-mobility short channel samples,
the coherence in the wave functions of electron pairs may be maintained during
pair transport along the channel. Additionally, if the transmission coefficient for
the injection of electron pairs into the channel is close to unity in the absence of
a significant reflection from potential barriers in their path and inelastic scattering between different pair states,
we are able to use a quantum ballistic transport model for interacting electron pairs
as far as the Coulomb interaction between electrons in the channel is fully taken
into account. Our quantum ballistic system with transport of interacting electron
pairs is shown schematically in Fig.\,\ref{FIG:0}.
\medskip

For a fixed linear electron density $n_{\rm 1D}$, the pair chemical potential
$\mu_{\rm p}(T,n_{\rm 1D})$  within the channel can be determined from

\begin{equation}
n_{\rm 1D}=\frac{2}{\pi}\,\sum_{j=1}^{4}\,\int\limits_{0}^\infty
dk_y\left\{\exp\left[\frac{E^{(p)}_{j,k_y}-\mu_{\rm
p}}{k_{\rm B}T}\right]+1\right\}^{-1}\ ,
\label{e2}
\end{equation}
where $k_y$ is the
wave vector of electrons along the channel, $T$ is the system temperature,
$E_{j,k_y}^{(p)}=E_{j}^{(p)}+\hbar^2k_y^2/m^\ast$, labeled by (p) for
$j=1, 2, 3, 4$, represents the lowest four conduction energy subbands of an interacting electron pair,
and $m^\ast$ is the effective electron mass.
In addition, the chemical potentials for the left and right electrodes are
$\mu_{\rm L}^{(p)}(V_b,\,T,n_{\rm 2D})=\mu_p(T,n_{\rm 1D})+eV_b$ and
$\mu_{\rm R}^{(p)}(V_b,\,T,n_{\rm 2D})=\mu_p(T,n_{\rm 1D})-eV_b$ in the presence
of the low biased voltage $V_b$, where $V_b$ is the applied biased voltage.
\medskip

For quantum ballistic charge/heat transport of interacting electron pairs in the channel,
the charge $(\alpha =0)$ and the heat $(\alpha =1)$ current densities are calculated
according to\,\cite{lyo5}

\begin{equation}
J^{(\alpha)}(V_b,\,T,\,n_{1D})=\frac{(-2e)^{1-\alpha}}{\pi}\,\sum_{j=1}^{4}\,\int\limits_{0}^\infty dk_y\,
(E_{j,k_y}^{(p)}-\mu_{\rm p})^\alpha\,
\left|v_{j,k_y}\right|
\left[f_{\rm L}(E^{(p)}_{j,k_y})-f_{\rm R}(E^{(p)}_{j,k_y})\right]\ ,
\label{e1}
\end{equation}
where $v_{j,k_y}=\hbar k_y /m^\ast$ is the
group velocity of an electron pair, $f_{\rm L}(E^{(p)}_{j,k_y})$ and $f_{\rm R}(E^{(p)}_{j,k_y})$ correspond to Fermi functions for noninteracting
electron pairs in the left (L) and right (R) electrodes with associated chemical potentials
$\mu_{\rm L}^{(p)}$	and $\mu_{\rm R}^{(p)}$	for noninteracting pairs, respectively.
\medskip

For the interacting electron pair, its energy levels  $E_{j}^{(p)}=E_{j,k_y=0}^{(p)}$,
as shown in Fig.\ \ref{FIG:1}, are calculated as $E_{1}^{(p)}\equiv E_{-}^{(p)}=\varepsilon_0+\varepsilon_1+
(u_{11}+u_{22})/2-\Delta_{\rm C}$, $E_{2}^{(p)}\equiv E_{+}^{(p)}=\varepsilon_0+\varepsilon_1+
(u_{11}+u_{22})/2+\Delta_{\rm C}$, $E_{3}^{(p)} =\varepsilon_0+\varepsilon_1+u_{33}$ and
$E_{4}^{(p)} =\varepsilon_0+\varepsilon_1+u_{44}$, where the Coulomb coupling term for the hybrid pair states is given by
$\Delta_{\rm C}=\sqrt{\left[\varepsilon_{1}-\varepsilon_{0}+(u_{22}-u_{11})/2\right]^2+|u_{12}|^2}$. Here, the
single-particle energy levels for harmonic-potential model with harmonic frequencies $\omega_x$
and $\omega_y$ in the transverse ($x$) and longitudinal ($y$) directions, respectively,
are $\varepsilon_0=(\hbar\omega_x+\hbar\omega_y)/2$ and
$\varepsilon_1=(3\hbar\omega_x+\hbar\omega_y)/2$, while the
employed Coulomb interaction energies are found to be
$u_{11}=N_0^2\,E_{\rm c}\,{\cal I}_{00,00}$, $u_{12}=N_0N_1\,E_{\rm c}\,{\cal I}_{00,11}$,
$u_{22}=N_1^2E_{\rm c}\,{\cal I}_{11,11}$, $u_{33}=N_0N_1\,E_{\rm C}\left({\cal I}_{01,01}+{\cal I}_{01,10} \right)$
and $u_{44}=N_0N_1\,E_{\rm c}\left(3\,{\cal I}_{01,01}-{\cal I}_{01,10}\right)$, where
$E_{\rm c}=e^2/4\pi\epsilon_0\epsilon_rL_y$ in terms of the length $L_y$ of the channel
and the background dielectric constant $\epsilon_r$, $N_n=\{\exp[(\varepsilon_n-\mu_0)/k_{\rm B}T]+1\}^{-1}$
($n=0,\,1$) is the single-particle level occupation factor, and $\mu_0(T,\,n_{\rm 1D})$ is the single-electron chemical potential.
Right before a pair of electrons is being injected into a conduction channel, these two electrons can select individual subband (same or different subbands and lower or higher subbands) for their ballistic transport.
Such a selection is subjected to subband population by the pool of electrons within the channel.
Right after this pair of electrons are injected into the channel, they will interact to each other through either intrasubband or intersubband Coulomb coupling.
The ballistic injection of electron pairs and the existence of an electron pool in the conduction channel are reflected in the inclusion of these two level occupation factors.
The symbol ${\cal I}_{\alpha\beta,\gamma\delta}$ represents the Coulomb integral for
$\alpha,\beta,\gamma,\delta=0,1$ if we only consider interacting pair states formed from
the lowest (`$0$') and first excited  (`$1$') state.
\medskip

Finally, for the harmonic-potential model, the four dimensionless Coulomb
integrals introduced above are calculated as

\begin{eqnarray}
{\cal I}_{00,00}({\cal R})&=&\sqrt{\frac{2}{\pi}}  \int^{\pi}_0
\frac{d\theta}{\left[1+({\cal R}^2-1)\cos^2\theta\right]^{1/2}}\ ,
\nonumber\\
{\cal I}_{11,11}({\cal R})  &=&\sqrt{\frac{2}{\pi}}  \int^{\pi}_0
\frac{d\theta}{\left[1+({\cal R} ^2-1)\cos^2\theta\right]^{1/2}}
\nonumber\\
&\times& \left\{1-\frac{{\cal R} ^2\cos^2\theta}{1+({\cal R} ^2-1)\cos^2\theta}+\frac{3{\cal R} ^4\cos^4\theta}{4\left[1+({\cal R} ^2-1)\cos^2\theta\right]^2}\right\}\ ,
\nonumber\\
{\cal I}_{01,01}({\cal R})  &=&   \frac{1}{\sqrt{2\pi}}   \int^{\pi}_0
d\theta\,\frac{2+({\cal R} ^2-2)\cos^2\theta}{\left[1+({\cal R} ^2-1)\cos^2\theta\right]^{3/2}}\ ,
\nonumber\\
{\cal I}_{01,10}({\cal R}) &=&    \frac{1}{\sqrt{2\pi}}     \int^{\pi}_0
d\theta\,\frac{{\cal R} ^2\cos^2\theta}{\left[1+({\cal R} ^2-1)\cos^2\theta\right]^{3/2}} \ ,
\label{e3}
\end{eqnarray}
where  the parameter ${\cal R}=W_x/L_y$ is the geometric ratio with $W_x$ denoting the width of the
conduction channel. For ${\cal R}\gg 1$, all the four terms in Eq.\,(\ref{e3}) scale as $1/{\cal R}$.
\medskip

By using the calculated $J^{(\alpha)}(V_b,\,T,\,n_{\rm 1D})$ in Eq.\,(\ref{e1}),
the electrical conductance $G(T,\,n_{\rm 1D})$ and the diffusion thermoelectric power $S_{\rm d}(T,\,n_{\rm 1D})$
of an interacting electron pair can be expressed as\,\cite{lyo5}

\begin{eqnarray}
G(T,\,n_{\rm 1D})&=&\frac{J^{(\alpha=0)}(V_b,\,T,\,n_{\rm 1D})}{V_b}\ ,
\nonumber\\
S_{\rm d}(T,\,n_{\rm 1D})&=&\frac{1}{T}\,\frac{J^{(\alpha=1)}(V_b,\,T,\,n_{\rm 1D})}{J^{(\alpha=0)}(V_b,\,T,\,n_{\rm 1D})}\ .
\label{e4}
\end{eqnarray}
In the next section, we present and discuss our numerical calculation and
their relationship to the recently reported results in Ref.\,[\onlinecite{pepper}].

\section{Discussion}
\label{sec:3}

\subsection{Theoretical Results}
\label{sec:3.1}

In all our numerical calculations, we set $T=10$\,mK, $V_b=0.01$\,mV, $L_y=400$\,nm,
$\epsilon_r=12$, and $m^\ast/m_0=0.067$ (with free-electron mass $m_0$).
Here, the quantum ballistic transport of hybrid pairs of electrons through a conduction
channel is defined as one moving through either one of orbit-triplet states $E^{(p)}_{\pm}$.
\medskip

\begin{figure}[t]
\centering
\includegraphics[width=0.65\textwidth]{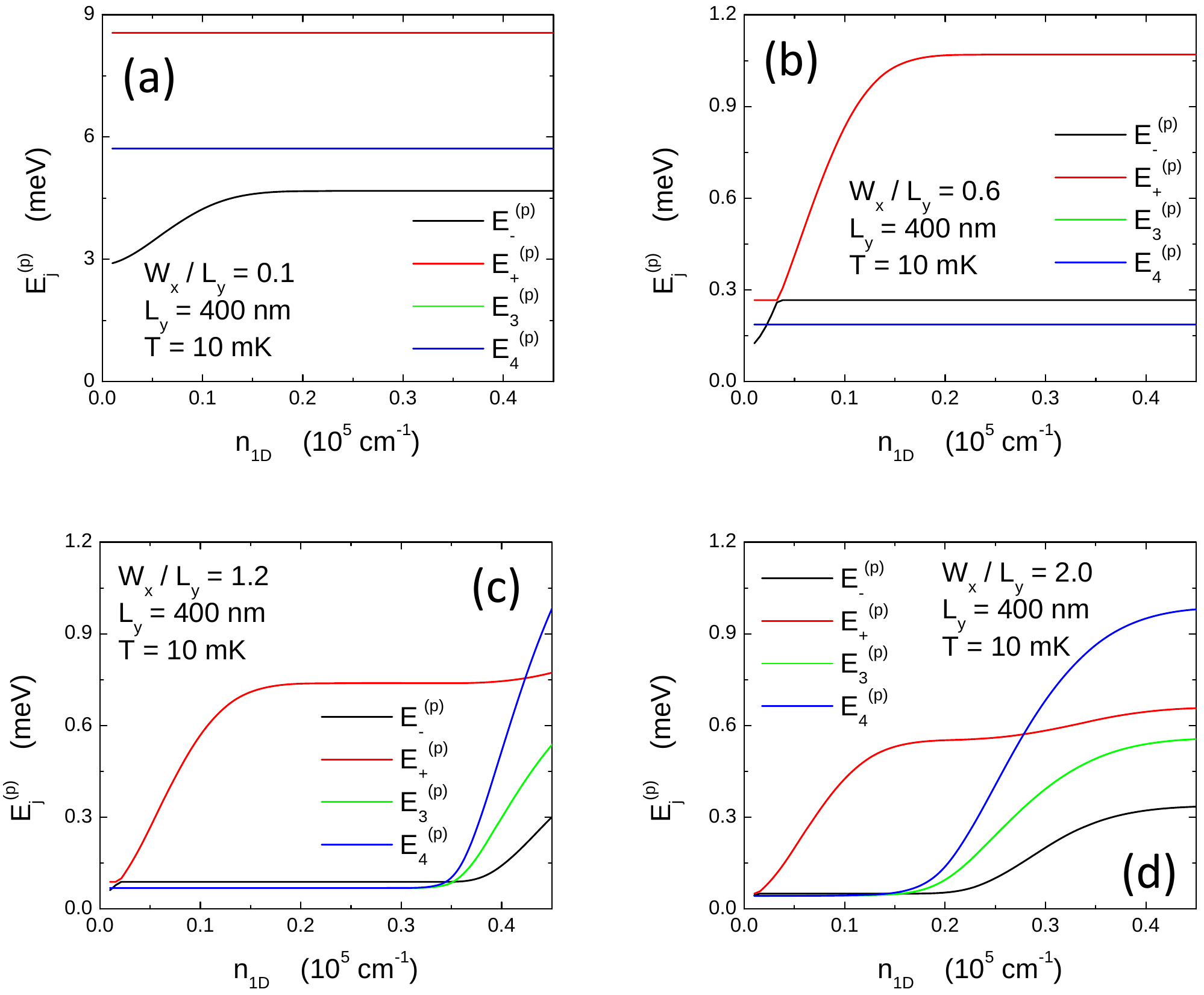}
\caption{(Color online) Plots of energy levels $E_j^{(p)}$ ($j=-,\,+,\,3,\,4$) of an interacting electron pair as a function of linear electron density $n_{\rm 1D}$ with several values of ${\cal R}=W_x/L_y$.
Here, we set ${\cal R}=0.1$ (a), $0.6$ (b), $1.2$ (c) and $2.0$ (d) for very strong to intermediate confinement.}
\label{FIG:1}
\end{figure}

For an interacting (hybrid) electron pair, their energy levels $E_j^{(p)}$ are expected
to depend on the electron density $n_{\rm 1D}$, as shown in Fig.\,\ref{FIG:1}. When the geometry ratio
${\cal R}=W_x/L_y$ is small for strong confinement in (a), only the ground state
$E^{(p)}_{-}$ is affected by varying  $n_{\rm 1D}$. As ${\cal R}$ increases to $0.6$ in (b), the level crossing between
$E^{(p)}_{-}$ of the hybrid state and the degenerate $E^{(p)}_3=E^{(p)}_4$ of the unhybrid state, as well as level anticrossing
between $E^{(p)}_{-}$ and $E^{(p)}_{+}$ of two hybrid states, occur at lower densities. When ${\cal R}>1$, as displayed in (c)
and (d), Coulomb interaction between electrons becomes significant. As a result, both $E^{(p)}_3$
and $E^{(p)}_4$ levels of two hybrid states are greatly pushed up at higher densities (i.e., $N_1>0$), leading to a recovery of the ground
state to $E^{(p)}_{-}$. In addition, the $E^{(p)}_4$ level for the spin-$1$ state in (c)
and (d) changes from the degenerate ground state at lower $n_{\rm 1D}$ to the highest-energy
state at higher $n_{\rm 1D}$. On the other hand, in the presence of a transverse magnetic field, the $E^{(p)}_3$ level for the spin-$0$ state
decouples from the magnetic field, while the degenerated $E^{(p)}_4$ level for the spin-$1$ state will
be split into three by the Zeeman effect,
leading to new $e^2/h$ and $3e^2/h$ conductance plateaus\,\cite{half}.
\medskip

\begin{figure}[t]
\centering
\includegraphics[width=0.65\textwidth]{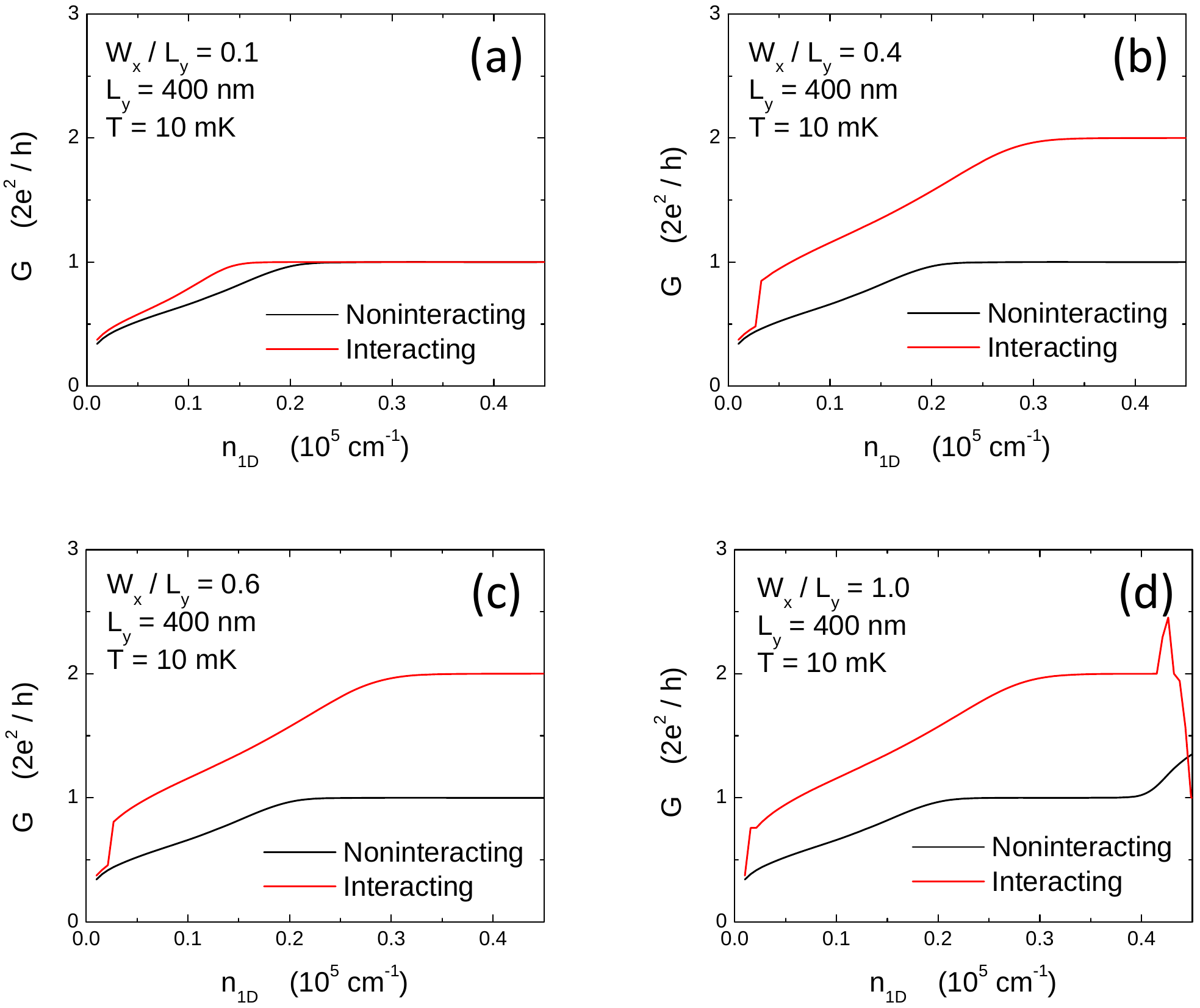}
\caption{(Color online) Plots of conductance $G$ as a function of $n_{\rm 1D}$ with several values of ${\cal R}$ for both noninteracting (black curves) and interacting (red curves) cases.
Here, we set ${\cal R}=0.1$ (a), $0.4$ (b), $0.6$ (c) and $1.0$ (d)  for intermediate to weak confinement.}
\label{FIG:2}
\end{figure}

Figure\ \ref{FIG:2} presents a comparison of conductance $G$ for both a non-interacting and an
interacting electron pair in the range of $0.1\leq {\cal R}\leq 1$. For very strong confinement in (a),
the Coulomb interaction effect is negligible and a conductance $2e^2/h$ plateau is clearly seen.
As ${\cal R}$ increases to $0.4$ in (b) and $0.6$ in (c) for cases with strong confinement,
$G$ for a non-interacting electron pair remains largely unchanged. For an interacting electron pair,
however, the conductance $2e^2/h$ plateau in (a) is completely destroyed by Coulomb interaction and accompanied by the occurrence
of a new $4e^2/h$ plateau for $G$. This behavior agrees with the result of both a unhybrid level-crossing and a hybrid
level anticrossing observed in Fig.\,\ref{FIG:1}(b).
This new $4e^2/h$ conductance plateau
is greatly perturbed at higher densities by a sharp spike and a follow-up deep dip to the lower $2e^2/h$ plateau
as ${\cal R}=1$ for intermediate confinement in (d).
\medskip

\begin{figure}[t]
\centering
\includegraphics[width=0.65\textwidth]{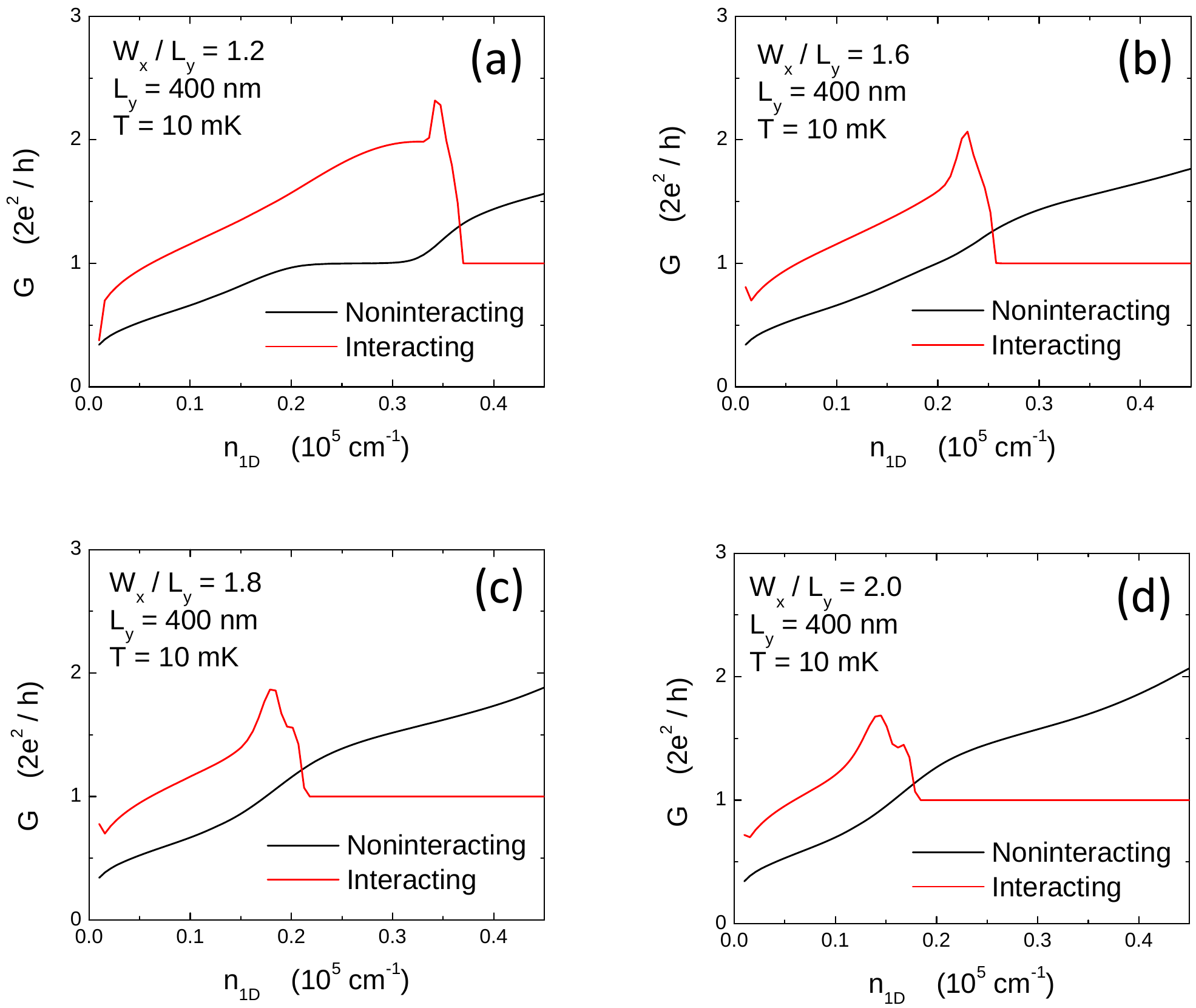}
\caption{(Color online) Plots of conductance $G$ as a function of $n_{\rm 1D}$ with several values of ${\cal R}=W_x/L_y$ for both noninteracting (black curves) and interacting (red curves) cases.
Here, we set ${\cal R}=1.2$ (a), $1.6$ (b), $1.8$ (c) and $2.0$ (d).}
\label{FIG:3}
\end{figure}

We present in Fig.\,\ref{FIG:3} the evolution of the conductance plateau with increasing ${\cal R}$ in the
weak confinement regime. When ${\cal R}\geq 1.6$, conductance plateaus for the non-interacting electron pair
are washed out in (b), (c) and (d) due to very small single-particle energy level separation in comparison
with the thermal energy $k_{\rm B}T$. It is also clear that the incomplete $4e^2/h$ conductance plateau in (a) for the
interacting electron pair is completely destroyed in this regime. However, the recovery of the single-particle-like $2e^2/h$
plateau, as seen in Fig.\,\ref{FIG:2}(a), can be seen in this plot. Additionally, the $2e^2/h$ plateau further
expands and extends to lower and lower electron densities
as ${\cal R}$ increases up to $2.0$. This unique reoccurrence feature can be fully explained by the
rising energy levels at higher densities due to relatively enhanced Coulomb repulsion as displayed in Figs.\,\ref{FIG:1}(c) and (d).
\medskip

For clarity, we note that as  a pair of  electrons are injected into a conduction channel, they may
 select specific subbands for their transport ballistically. This selection rule  is determined
by  the occupation factor of the electrons already within the channel. During the period of time
that the two injected moving electrons are within the channel, they may interact with each other through
either the intrasubband or the intersubband Coulomb coupling. We emphasize that the linear density
of electrons confined within the channel may be held constant when the channel width is varied.
For this to occur, the Fermi energy will adjust itself to accommodate all electrons and additional
subbands are populated accompanied by  reduced energy level separations. Specifically, the Fermi
energy is actually reduced and the number of electrons in the channel is not changed at all.
Furthermore, although the Fermi energy is reduced, the second level may still be populated due
to reduced level separation to keep the number of electrons in the channel a constant.
Clearly,    enhancement of the  Coulomb interaction is not solely determined  by the electron density,
since it also depends on how electrons are distributed.  For the Coulomb effect on the states of
the pair  of electrons, the inclusion of a new populated pair state, with one electron in a lower energy level
and the  other electron in a higher level, will  induce a new Coulomb effect on the pair states of electrons.
\medskip

\begin{figure}[t]
\centering
\includegraphics[width=0.65\textwidth]{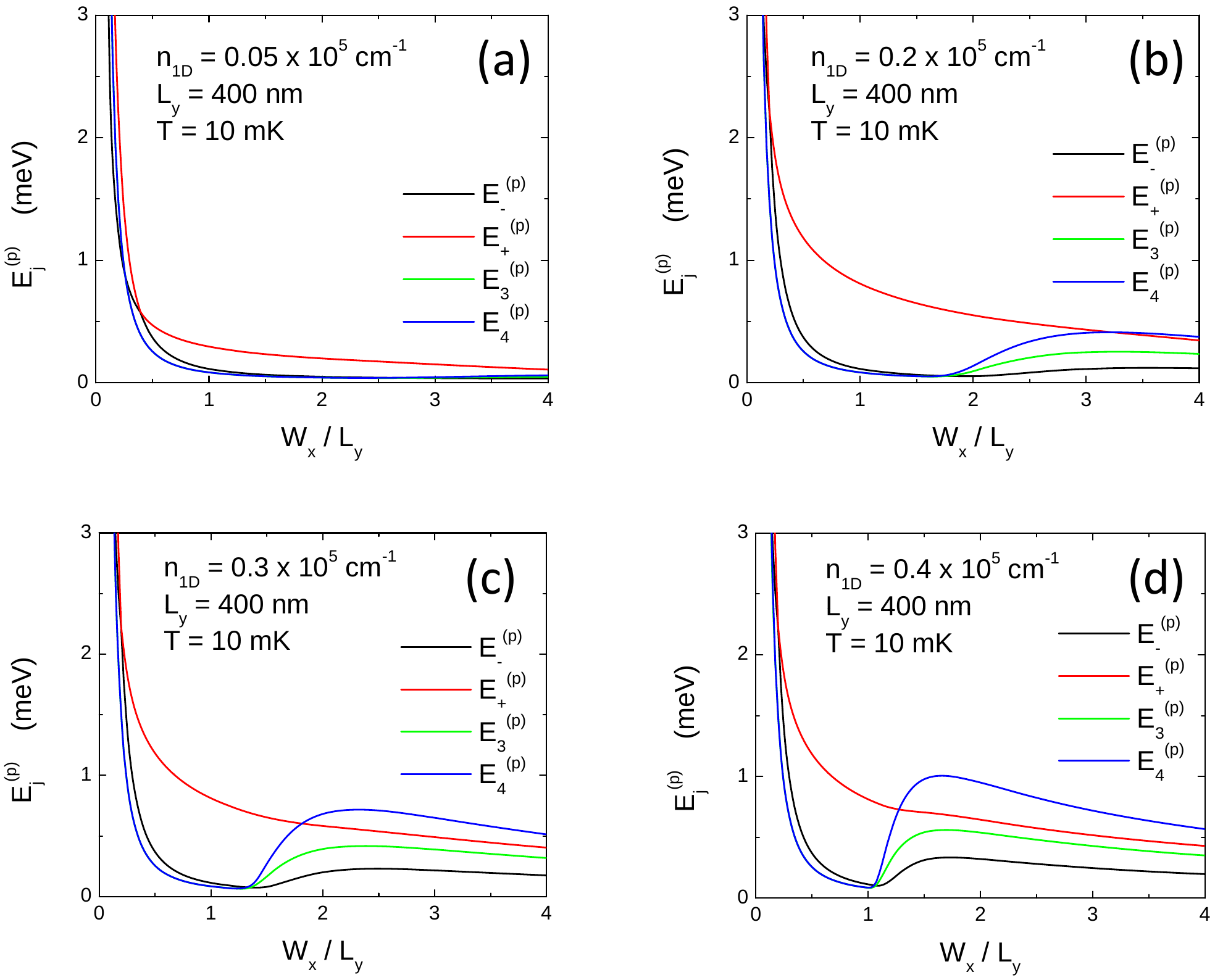}
\caption{(Color online) Plots of energy levels $E_j^{(p)}$ ($j=-,\,+,\,3,\,4$) of an interacting electron pair as a function of ${\cal R}$ with several values of $n_{\rm 1D}$.
Here, we set $n_{\rm 1D}=0.05$ (a), $0.2$ (b), $0.3$ (c) and $0.4$ (d) in unit of $10^{5}$\,cm$^{-1}$ for increasing the significance of Coulomb interaction.}
\label{FIG:4}
\end{figure}

As the transverse confinement becomes weaker (or the ${\cal R}$ value is increased), the
kinetic part of the energy levels $E_j^{(p)}$ of a pair will drop as $1/{\cal R}^2$ for fixed $L_y$.
Therefore, by increasing ${\cal R}$, the significance of Coulomb interaction, which scales
as $1/{\cal R}$ as shown by Eq.\,(\ref{e3}), will be relatively enhanced. In addition, the
second energy level will be occupied by increasing ${\cal R}$ for fixed electron density due
to reduced level separation. Consequently, the additional Coulomb repulsion between two
electrons on different energy levels is introduced. This effect is reflected in
Figs.\,\ref{FIG:4}(b), (c) and (d) as pushing up the energy levels $E^{(p)}_-$ and $E^{(p)}_3$
(as $N_1>0$) in the region of ${\cal R}>1$ when $n_{\rm 1D}\geq 0.2\times 10^{5}$\,cm$^{-1}$.
Furthermore, the existence of three-fold spin degeneracy in the $E^{(p)}_4$ level pushes
itself above the $E^{(p)}_3$ level as the Coulomb interaction is enhanced for ${\cal R}>1$.
However, the $E^{(p)}_+$ electron pair state, associated with two excited-state electrons,
is still dominated by the kinetic energy for the whole range of ${\cal R}$ shown in this figure.
As $n_{\rm 1D}$ further increases, the Coulomb repulsion effect extends to the intermediate
confinement regime in Fig.\,\ref{FIG:4}(d). As a whole, we find the ground state $E^{(p)}_-$ level
in (a) for small values of ${\cal R}$ and $n_{\rm 1D}$ (where
the kinetic energy of electrons is dominant) is fully recovered in (d) for large values of ${\cal R}$
and $n_{\rm 1D}$ (where the Coulomb energy is dominant).
It is interesting to note that there exists an intermediate confinement regime
(${\cal R}\gtrapprox 1$) between the strong (scaling as fast drop $1/{\cal R}^2$ for ${\cal R}<1$)
and weak (scaling as slow drop $1/{\cal R}$ for ${\cal R}\gg 1$)
confinement regimes, where the Coulomb interaction between electrons can be relatively highlighted to give rise to pushing up of three energy levels and the recovery of the the ground state $E^{(p)}_-$ level simultaneously.
\medskip

\begin{figure}[t]
\centering
\includegraphics[width=0.65\textwidth]{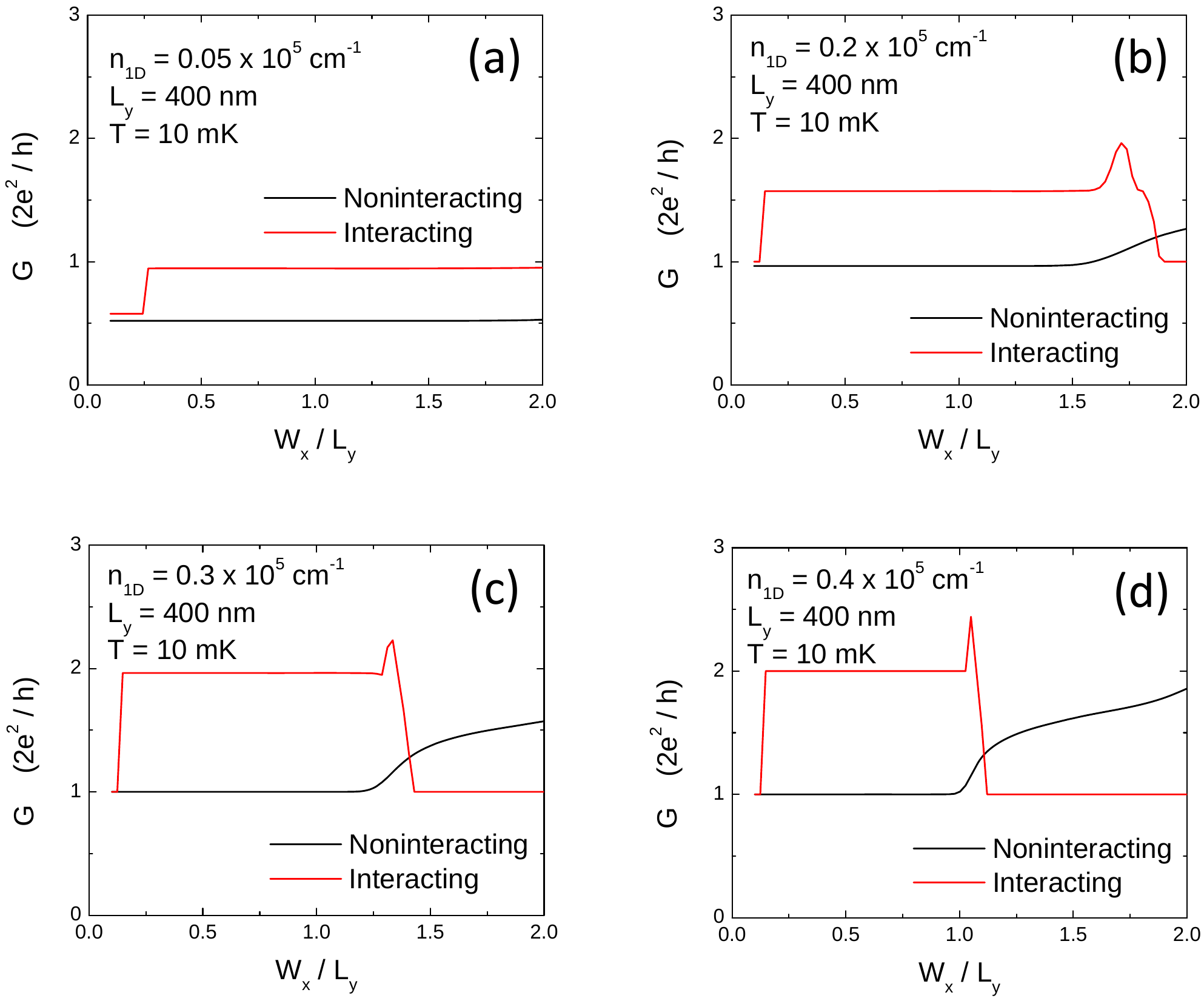}
\caption{(Color online) Plots of conductance $G$ as a function of ${\cal R}$ with several values of $n_{\rm 1D}$ for both noninteracting (black curves) and interacting (red curves) cases.
Here, we set $n_{\rm 1D}=0.05$ (a), $0.2$ (b), $0.3$ (c) and $0.4$ (d) in unit of $10^{5}$\,cm$^{-1}$.}
\label{FIG:5}
\end{figure}

The ground-state recovery observed in Fig.\,\ref{FIG:4} has a profound influence both on the distribution of
conductance plateaus and on the interplay of the electron hybridization, as displayed in Fig.\,\ref{FIG:5}.
When $n_{\rm 1D}$ is very small, the Coulomb effect can be neglected. In this case, the $2e^2/h$
conductance plateau is observed for the interacting electron pair as shown in (a) for all values of ${\cal R}$. As $n_{\rm 1D}$ is
increased to $0.2\times 10^{5}$\,cm$^{-1}$ in (b), the $2e^2/h$ plateau in (a) is destroyed except
for its recovery close to ${\cal R}=2.0$. If the value of $n_{\rm 1D}$ is further increased as
in (c) and (d), the new $4e^2/h$ conductance plateau shows up for the interacting electron pair,
which corresponds to the population of the degenerated lowest energy levels $E^{(p)}_3$ and $E^{(p)}_4$ after their crossing another $E^{(p)}_-$ energy level.
As ${\cal R}$ further increases above one in the very-weak confinement regime, the ground-state recovery, as discussed in Figs.\,\ref{FIG:4}(c) and (d), enforces
the reoccurrence of the $2e^2/h$ conductance plateau due to strong Coulomb
repulsion between electrons.
\medskip

\begin{figure}[t]
\centering
\includegraphics[width=0.65\textwidth]{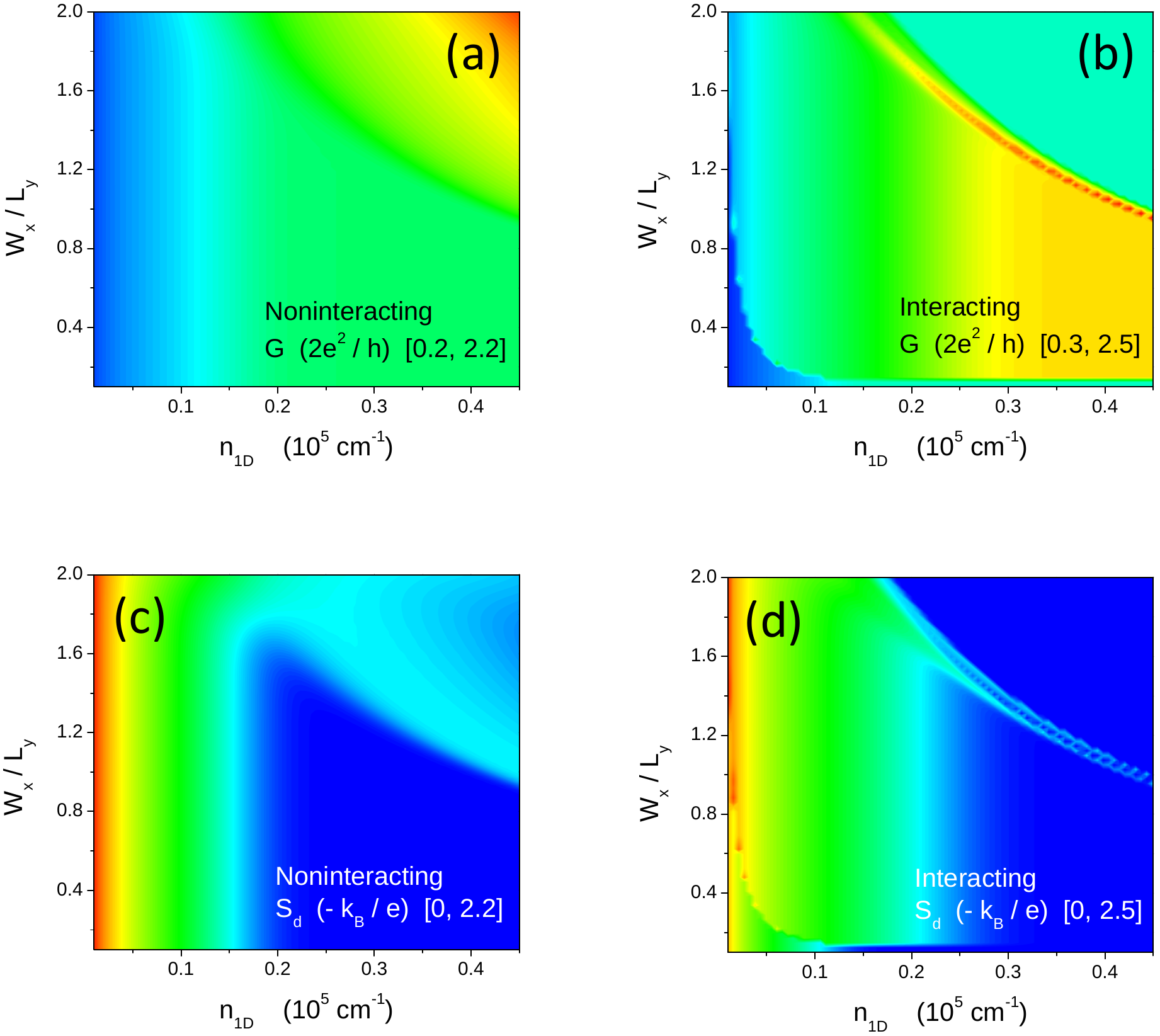}
\caption{(Color online) Contour plots of $G$ [(a), (b)] and $S_{\rm d}$ [(c), (d)] as functions of both $n_{\rm 1D}$ and ${\cal R}$ for either noninteracting [(a), (c)] or interacting [(b), (d)] case.}
\label{FIG:6}
\end{figure}

In order to get a complete picture of the quantum ballistic transport of interacting pairs of electrons passing through
a one-dimensional conduction channel, we present the contour plots of conductance $G$ and diffusion thermoelectric
power of electrons $S_{\rm d}$ as functions of ${\cal R}$ and $n_{\rm 1D}$ in Fig.\,\ref{FIG:6} for both
non-interacting and interacting electron pairs as a comparison. By comparing (a) and (b) for $G$, we find that the Coulomb effect is
most dominant in the upper right-hand corner region of (b) within a weak confinement regime and a relatively high electron
density, where a gradual conductance is replaced by a $2e^2/h$ conductance plateau due to strong Coulomb repulsion between electrons. In addition, we
also find another $4e^2/h$ conductance plateau in the lower-right corner region of (b) (which is separated by a spike in $G$
from the upper-right corner region), where confinement is intermediate or stronger but the electron density is high.
From the comparison of (c) and (d), we find that the Coulomb interaction suppresses $S_{\rm d}$ in the weak
confinement region and with a relatively high electron density.
Under very strong confinement, a downward step in $S_{\rm d}$ outside its suppression region is seen for interacting electron pairs.
Moreover, the spike in $G$ also has a visible feature reflected in $S_{\rm d}$.

\subsection{Experimental Verification}
\label{sec:3.2}

Two-terminal differential conductance measurements were performed
using an excitation voltage of $10\,\mu$V at $73$\,Hz using the Oxford
Instruments cryofree dilution refrigerator, where the device was
estimated to have an electron temperature of around $70$\,mK.
\medskip

A top gated, split gate device provides additional confinement to the
1D electrons which enables varying the confinement from being very
strong (zero top gate) to very weak (very negative top gate voltage).
In the present work, top gate voltage, $V_{\rm tg}$ was varied from $0$ (left) to
$-2.2$\,V (right) in the steps of $50$\,mV.
\medskip

The device used in the present work was fabricated from a modulation doped GaAs/AlGaAs heterostructure grown using a molecular beam epitaxy (MBE), where a two-dimensional electron gas (2DEG)
is formed $300$\,nm beneath the interface.
Typical dimensions of the split gate device are: length $400$\,nm and width $700$\,nm. A top gate covers the entire split gate sandwiching a crossed-linked PMMA layer of thickness $200$\,nm.
The 2DEG sits around $300$\,nm beneath the surface of the GaAs/AlGaAs heterostructure. The 2DEG mobilities and electron densities are $3.5$-$5.0\times 10^6$\,cm$^2$/Vs and $1.8$-$2.2\times 10^{11}$\,cm$^{-2}$,
respectively.
\medskip

\begin{figure}[t]
\centering
\includegraphics[width=0.65\textwidth]{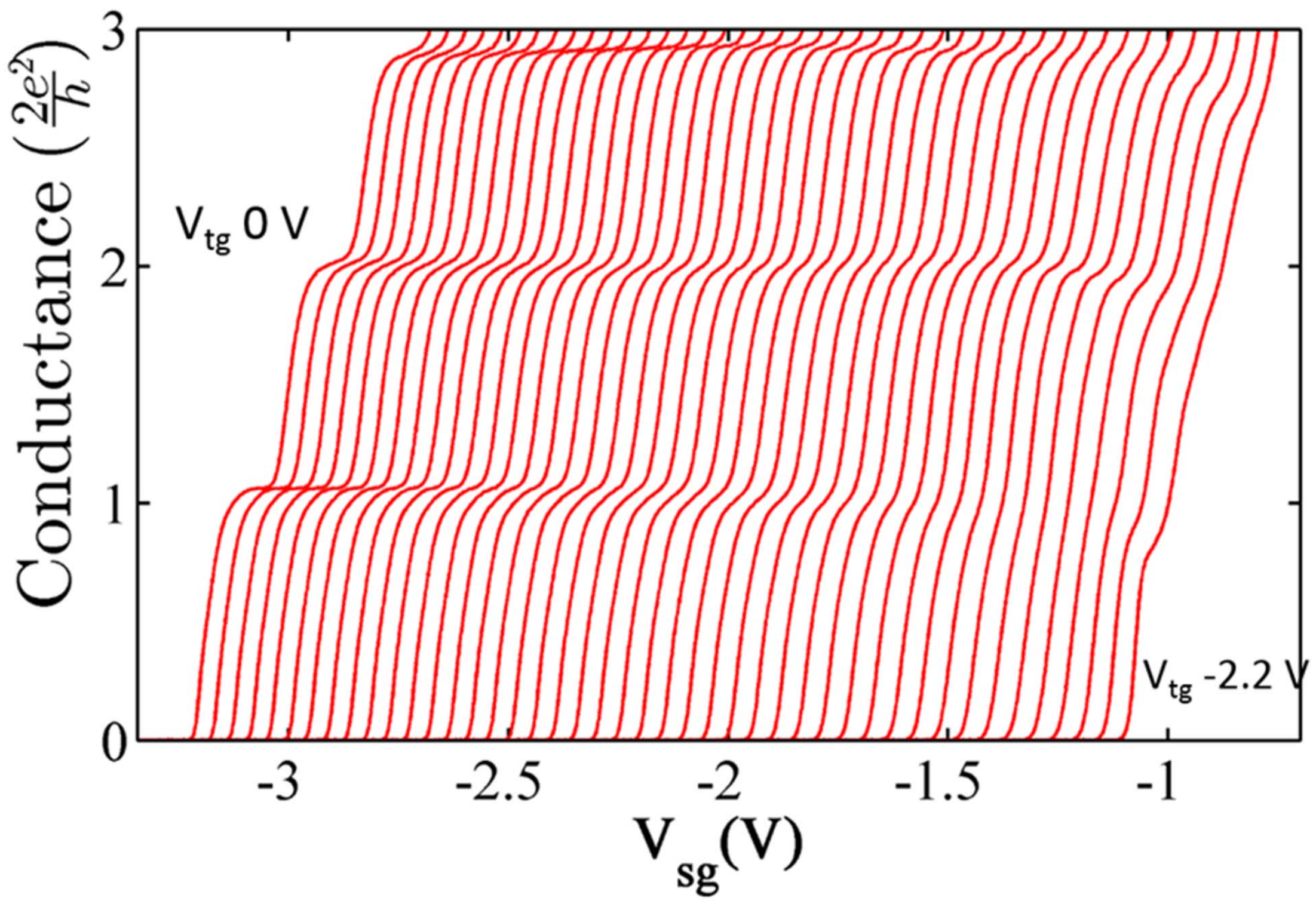}
\caption{(Color online) Plot for measured differential conductance, where a jump to $4e^2/h$ occurs when the confinement is weakened using a top gated,
split-gate device. The confinement is controlled by making the top gate negative so that left(right) of the plot is strong(weak) confinement.}
\label{FIG:7}
\end{figure}

Figure\ \ref{FIG:7} shows the differential conductance plot of the device as a
function of split gate voltage for various top gate voltages.
As shown in Fig.\,\ref{FIG:7}, when the confinement is weakened, the $2e^2/h$ conductance plateau weakens. On further weakening the confinement, the $2e^2/h$
plateau disappears and a direct jump in conductance to the $4e^2/h$ plateau occurs at $V_{\rm tg}=-2.05$\,V. Eventually the first plateau at $2e^2/h$
comes back on further weakening the confinement at $V_{\rm tg}=-2.2$\,V.
\medskip

On the other hand, from our calculated results in Figs.\,\,\ref{FIG:2} and \,\ref{FIG:3},
we see the occurrence of the $2e^2/h$ conductance plateau for small values
of ${\cal R}$ in the strong-confinement regime, the $4e^2/h$ conductance plateau
for intermediate confinement, and the $2e^2/h$ conductance plateau
in the weak-confinement regime preceded by a double-kink structure.
\medskip

Therefore, we conclude from above that the experimental observations agree well with our theoretical prediction in this paper.
Therefore, this experimentally observed feature for switching conductance plateau can be explained
by the switching of the ground state from $E^{(p)}_-$ to degenerated $E^{(p)}_3$ and $E^{(p)}_4$ and back to $E^{(p)}_-$, which is reflected as an
upward jump from $2e^2/h$ to $4e^2/h$ and followed by another downward jump from $4e^2/h$ back to
$2e^2/h$ with increasing channel width.

\section{Concluding Remarks}
\label{sec:4}

The ballistic conductance for a quasi-one dimensional channel
(quantum wire) has exhibited interesting behavior as functions
of the electron density as well as the confinement.
We theoretically demonstrated that the electron-electron
interaction explicitly plays a crucial role in our calculations in a weak-confinement regime.
We  carried out an extensive calculation of the effect of
confinement on the conductance and the associated dependence
on the interplay of hybrid and unhybrid quantum transport of two electrons.
As shown through our numerical calculations, depending on the
confinement parameter the conductance manifests the signature
of single particle or hybrid particles behavior. This
dependence can be observed in the variation of the conductance
from $2e^2/h$ (single-particle) to $4e^2/h$ (unhybrid interaction) and back to $2e^2/h$ (hybrid interaction)
as a function of the width of the quantum wire. It is interesting to observe how many-body
effects enter into the calculation of the quantum ballistic conductance.

\newpage

\end{document}